\documentclass{aa}

\usepackage{graphicx}
\usepackage{txfonts}
\usepackage[draft]{hyperref}
\usepackage{savesym}
\usepackage{txfonts}
\usepackage{amsmath}
\usepackage{listings}             
\usepackage{color}
\definecolor{deepblue}{rgb}{0,0,0.5}
\definecolor{deepred}{rgb}{0.6,0,0}
\definecolor{deepgreen}{rgb}{0,0.5,0}
\definecolor{mygrey}{rgb}{0.5,0.5,0.5}

\usepackage{natbib,twoopt}
\bibpunct{(}{)}{;}{a}{}{,} 

\makeatletter
\newcommandtwoopt{\citeads}[3][][]{\href{http://adsabs.harvard.edu/abs/#3}%
{\def\hyper@linkstart##1##2{}%
\let\hyper@linkend\@empty\citealp[#1][#2]{#3}}}
\newcommandtwoopt{\citepads}[3][][]{\href{http://adsabs.harvard.edu/abs/#3}%
{\def\hyper@linkstart##1##2{}%
\let\hyper@linkend\@empty\citep[#1][#2]{#3}}}
\newcommandtwoopt{\citetads}[3][][]{\href{http://adsabs.harvard.edu/abs/#3}%
{\def\hyper@linkstart##1##2{}%
\let\hyper@linkend\@empty\citet[#1][#2]{#3}}}
\newcommandtwoopt{\citeyearads}[3][][]%
{\href{http://adsabs.harvard.edu/abs/#3}
{\def\hyper@linkstart##1##2{}%
\let\hyper@linkend\@empty\citeyear[#1][#2]{#3}}}
\makeatother

\begin{document}
\title{Cygrid: A fast Cython-powered convolution-based gridding module for Python}
\author{B. Winkel\inst{1}
        \and
        D. Lenz\inst{2}
        \and
        L. Fl\"{o}er\inst{2}
       }

\institute{Max-Planck-Institut f\"{u}r Radioastronomie (MPIfR),
              Auf dem H\"{u}gel\,69, 53121 Bonn, Germany\\
              \email{bwinkel@mpifr.de}
            \and
            Argelander-Institut f\"{u}r Astronomie (AIfA),
              Auf dem H\"{u}gel\,71, 53121 Bonn, Germany
            }

\date{Received ; accepted }

\abstract
{Data gridding is a common task in astronomy and many other science disciplines. It refers to the resampling of irregularly sampled data to a regular grid.}
{We present cygrid\thanks{\url{https://github.com/bwinkel/cygrid}}, a library module for the general purpose programming language Python. Cygrid can be used to resample data to any collection of target coordinates, although its typical application involves FITS maps or data cubes. The FITS world coordinate system standard is supported.}
{The regridding algorithm is based on the convolution of the original samples with a kernel of arbitrary shape. We introduce a lookup table scheme that allows us to parallelize the gridding and combine it with the HEALPix tessellation of the sphere for fast neighbor searches.}
{We show that for $n$ input data points, cygrids runtime scales between $O(n)$ and $O(n\log n)$ and analyze the performance gain that is achieved using multiple CPU cores. We also compare the gridding speed with other techniques, such as nearest-neighbor, and linear and cubic spline interpolation.}
{Cygrid is a very fast and versatile gridding library that significantly outperforms other third-party Python modules, such as the linear and cubic spline interpolation provided by SciPy.}

\titlerunning{Cygrid: A fast gridding module for Python}
\authorrunning{Winkel, Fl\"{o}er, \& Lenz}

\maketitle

\section{Introduction}\label{sec:intro}
In many natural sciences, observational data are often irregularly sampled. For example, in astronomy a single-dish telescope actively scans the sky to produce a map. Even if the scan pattern was designed well, the measured data point locations might deviate from a perfect rectangular grid, for example, caused by pointing inaccuracies, the spherical geometry of the sky, or distortions in the optics. However, a regular grid is needed to visualize the data on pixel-based devices, enable algorithmic analysis, or store the data in FITS images. The task of resampling a data set from irregular spacing to a regular grid is called gridding.

There are already many different gridding techniques known in the literature. Some are easy to understand and implement, such as nearest-neighbor resampling, or linear and cubic spline interpolation, others involve complex statistical inference, for example, the Kriging algorithm \citep{krige51,matheron63}. Despite the fact that many available methods are fast and memory efficient, they are not always the best choice for astronomical applications because they do not conserve the flux density of a source.

There is a sufficiently fast alternative, convolution-based gridding, which is well known in many disciplines, especially in radio astronomy. For example, it is included in the long-known and mature interferometry software package AIPS\footnote{\url{http://www.aips.nrao.edu/index.shtml}} and its successor CASA\footnote{\url{http://casa.nrao.edu/}} (\texttt{sdgrid} task). It is also used by the \textsc{Gildas} software\footnote{\url{http://www.iram.fr/IRAMFR/GILDAS/}}. Furthermore, \citet{kalberla05,kalberla10} apply this gridding technique to produce (all-sky) \ion{H}{i} data cubes and column density maps for the Leiden/Argentine/Bonn Survey \citep[LAB;][]{kalberla05} and the Galactic All-Sky Survey \citep[GASS;][]{mcclure09}. We briefly recapitulate this algorithm in Section~\ref{subsec:basicgridding}.

As long as the applied gridding kernel is sampled with sufficient spatial density on the target grid, convolution-based gridding conserves flux densities. The algorithm can easily be implemented to work serially, which is beneficial for its memory footprint. To parallelize this scheme, we need to make sure that two threads never write to the same output element. In Section~\ref{subsec:advancedgridding}, we develop an efficient lookup scheme, which is based on the HEALPix tessellation of the sphere \citep{gorski05}, to optimally distribute the work across different threads. Another advantage of our proposed HEALPix lookup scheme is that it brings the time complexity close to $O(n),$ where $n$ is the number of input samples even for arbitrary world coordinate system \citep[WCS;][]{greisen02,calabretta02} projections, which is otherwise not easily possible. In Section~\ref{sec:cygrid} we discuss our reference implementation for the Python\footnote{\url{http://www.python.org/}} programming language, the \texttt{cygrid} module. It was developed in the framework of the Effelsberg--Bonn \ion{H}{i} Survey \citep[EBHIS;][]{kerp11,winkel16}. Cygrid makes use of Cython\footnote{\url{http://cython.org/}} to improve the speed of some performance-critical functions. Section~\ref{sec:benchmarks} presents speed tests and benchmarks. We conclude with a summary in Section~\ref{sec:summary}.

\section{The gridding algorithm}\label{sec:griddingalgorithm}

In astronomy, gridding tasks often have to contend with (spherical) sky coordinates.  The FITS WCS standard and its various software implementations are most convenient for the conversion between world coordinates and pixel grids. In the following, we constrain ourselves to the astronomy-specific case of gridding into WCS pixel grids.

\subsection{Basic method}\label{subsec:basicgridding}

The basic idea of convolution-based gridding is to compute the discrete convolution of the input data with a gridding kernel function. This is equivalent to calculate for each output grid point the weighted sum of relevant input data samples, where the weighting function is the gridding kernel,
\begin{equation}
R_{i,j}[s]=\sum_n R_n [s](\alpha_n,\delta_n)w(\alpha_{i,j},\delta_{i,j};\alpha_n,\delta_n)\,.
\label{eq:basicgriddingunweighted}
\end{equation}
Here, $R_n [s]$ and $R_{i,j}[s]$ are two different representations of the true signal $s$. The index $n$ runs over the list of all input samples, with respective coordinates $(\alpha_n,\delta_n)$, while the regular output grid can be parametrized via pixel coordinates $(i,j)$ with associated world coordinates $(\alpha_{i,j}, \delta_{i,j})$. The value of the weighting kernel $w$ depends only on the input and output coordinates. In most cases a radially symmetric kernel is applied, such that $w(\alpha_{i,j},\delta_{i,j};\alpha_n,\delta_n)=w\left(d(\alpha_{i,j},\delta_{i,j};\alpha_n,\delta_n)\right)$, with the true angular distance $d$. The true angular distance can be calculated with the Haversine formula or the Vincenty formula.\footnote{\url{https://en.wikipedia.org/wiki/Great-circle_distance}} The latter has better numerical accuracy for large angular separations, but is slightly more computationally demanding. For the gridding task, which involves calculation of small distances, the Haversine formula is sufficient.

If one is interested in flux density conservation, one has to account for the potential irregular sampling of the input data; different target pixels could be influenced by a very different number of input data samples. To account for this, we introduce the overall weight map,

\begin{equation}
W_{i,j}\equiv\sum_n w(\alpha_{i,j},\delta_{i,j};\alpha_n,\delta_n)\,,
\label{eq:weightmap}
\end{equation}
such that Eq.~(\ref{eq:basicgriddingunweighted}) becomes
\begin{equation}
R_{i,j}[s]=\frac{1}{W_{i,j}}\sum_n R_n[s](\alpha_n,\delta_n)w(\alpha_{i,j},\delta_{i,j};\alpha_n,\delta_n)\,.
\label{eq:basicgridding}
\end{equation}
Even with unnormalized gridding kernels, flux density conservation is then guaranteed.

To compute a complete map $R_{i,j}[s]$, one needs to iterate over all pixels $(i,j)$ and apply Eq.~(\ref{eq:basicgridding}) to each of them. In practice, it is easier to perform the outer iteration over $n$ instead because the input data set is usually larger than the output map. The following algorithm is then applied:

\begin{enumerate}
\item Set $R_{i,j}\equiv0$ and $W_{i,j}\equiv0$.
\item Iterate over all input samples $n$ and each target pixel $(i,j)$:
  \begin{itemize}
  \item calculate weight, $w$, for each coordinate pair $(\alpha_{i,j},\delta_{i,j}) \leftrightarrow (\alpha_n,\delta_n);$
  \item compute $R_n \cdot w$ , add the result to $R_{i,j}$, and add $w$ to $W_{i,j}$.
  \end{itemize}
\item Divide $R_{i,j}$ by $W_{i,j}$ pixel-wise to get the final map.
\end{enumerate}

\begin{figure}[!tp]
\centering%
\includegraphics[width=0.49\textwidth,viewport=40 40 392 392,clip=]{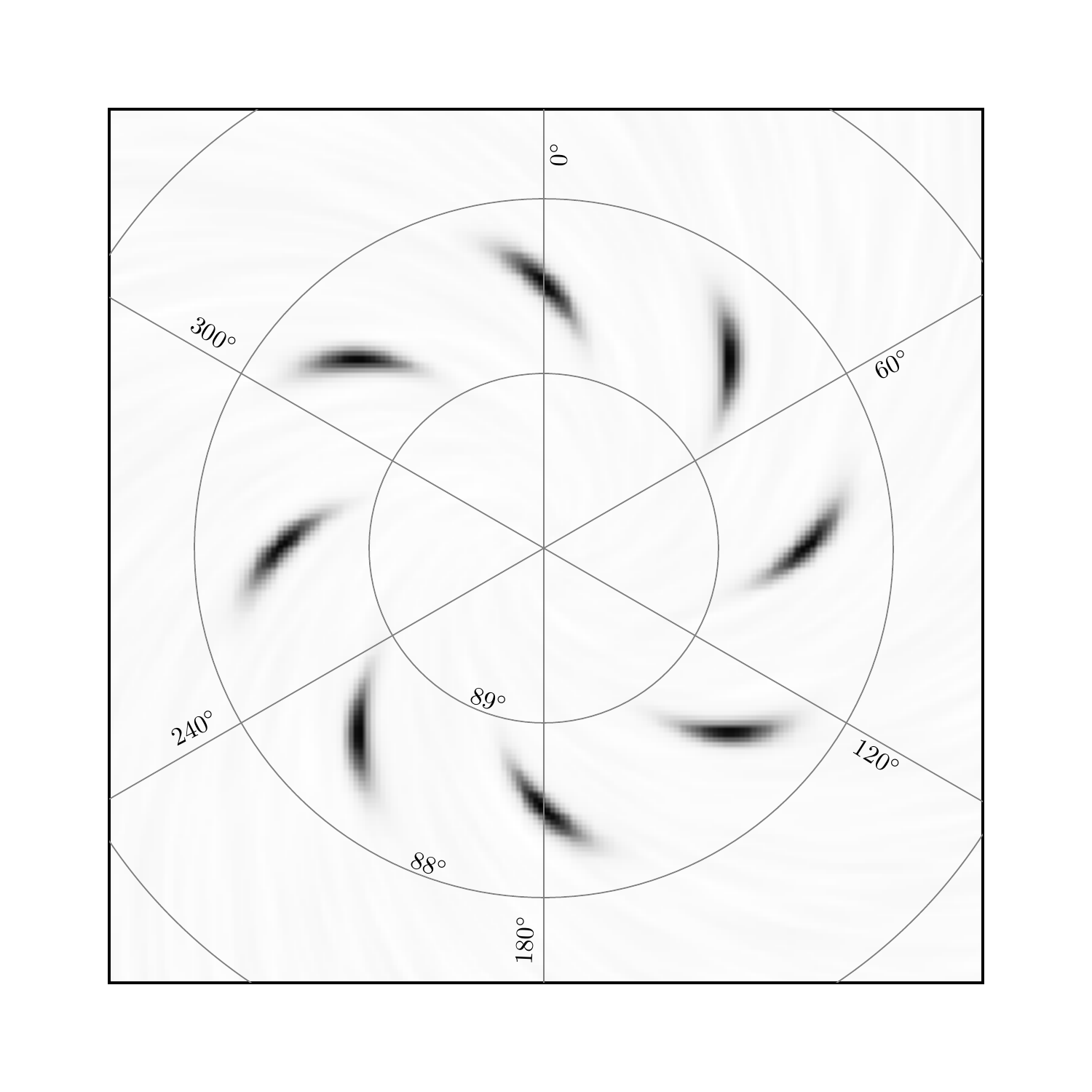}
\caption{Convolution-based gridding in the world coordinate system has the advantage that even relatively complicated tasks are easy to implement. In the visualized example an elliptical 2D-Gaussian gridding kernel with a bearing of $45\degr$ is used for a field close to the pole. Artificial point sources were inserted in intervals of $45\degr$ longitude at a latitude of $88.5\degr$. The resulting ellipses in the pixel space are distorted from the chosen projection (zenith-equal-area) of the sphere onto a plane.}%
\label{fig:expample_elliptical}%
\end{figure}

The presented method has several advantages. It is straightforward to implement and incorporate any WCS projection. By calculating the world coordinates of the target pixel grid, the method can convolve in world coordinate space rather than in pixel space. This also facilitates its application in more complicated tasks, for example, the use of elliptical Gaussian kernels in true world coordinates; see Fig.~\ref{fig:expample_elliptical}.

A disadvantage of the sum in Eq.~(\ref{eq:basicgridding}) is that it is not easy to parallelize the code. While one could easily compute partial sums simultaneously, the result of each atom $R_n \cdot w$ must be written into the target map $R_{i,j}$ and weight map $W_{i,j}$. In practice, this can lead to race conditions, when two threads try to write into the same memory cell simultaneously, which requires to employ a locking mechanism at the cost of processing speed. One can circumvent this problem by letting each thread use its own copy of $R_{i,j}$ and $W_{i,j}$, and merging them afterward. However, this would significantly increase the memory consumption for data cubes or very large target maps, even if only a small number of threads is involved.

Furthermore, the workload can greatly be reduced by preselecting input samples in the vicinity of each target pixel: all pixels with negligible  weight (for a given input sample) can safely be ignored. For some WCS projections this would be relatively easy, but others have discontinuities, which makes the neighbor search computationally more complex. Without such optimization the convolution had to be calculated for the full target grid size. This would make the algorithm $O(i\cdot j\cdot n)$ instead of $O(n)$, and results in a large number of unnecessary operations if the convolving kernel is much smaller than the size of the map.

\subsection{Improvement with HEALPix-based lookup tables}\label{subsec:advancedgridding}

We can avoid both of the previously mentioned disadvantages, if we introduce a lookup table based approach. This is best accomplished by internally working on the HEALPix grid. The HEALPix grid is a tessellation of the sphere, where each pixel has the same solid angle but the pixels are generally not rectangular. For information about HEALPix, see \citet{gorski05}.

There exist several software libraries related to HEALPix\footnote{e.g., \url{http://healpix.sourceforge.net/}}, which provide fast lookup routines to select HEALPix elements based on WCS coordinates. For example, a function is provided to obtain all the HEALPix pixels within a circle of given radius for a certain sky position, the so-called \texttt{query-disk} function. This can be used to greatly improve the complexity of the gridder from $O(i\cdot j\cdot n)$ to $O(n)$, i.e., to avoid the problem described in the last paragraph of Section~\ref{subsec:basicgridding}.

In the following, the basic concept of the improved algorithm is presented, while in Section~\ref{sec:cygrid} we discuss our reference implementation cygrid in more detail, along with some techniques to improve computational efficiency.

The main idea is to use a lookup table that contains, for each target pixel $(i,j),$ the subset of input sample indices $\{n\}_{i,j}$ that will significantly contribute to the convolution. Then, instead of having the outer loop iterate over $n$, we iterate over the list, $\left\{\{n\}_{i,j}\right\}$, of all non-empty $\{n\}_{i,j}$ and their respective elements. As such the outer loop can be computed concurrently because the program writes into different map and weight-map cells in each thread.

To determine the list $\left\{\{n\}_{i,j}\right\}$ several steps are necessary as follows:

\begin{enumerate}
\item Set the pixel resolution for the internally used HEALPix grid. In the HEALPix scheme, each coordinate pair is represented by a single index, which makes it easy to use the HEALPix index as hash table key. In Section~\ref{subsec:cygrid_example} we discuss how to find appropriate values.
\item For each target map pixel $(m, n)$ find the nearest HEALPix pixel $h_{i,j}$. In the resulting list $\{h_{i,j}\}$, the HEALPix indices are not necessarily unique because multiple input pixels could be associated with the same HEALPix pixel.
\item Iterate over all input sample indices, $n$
\begin{enumerate}
\item Calculate the associated HEALPix index, $h_n$.
\item Run the \texttt{query-disk} function to get all HEALPix indices $\{h_n^r\}$ within a radius of $r_d$ around $(\alpha_n,\delta_n)$. This radius is chosen such that the kernel has effectively zero contribution for larger angular separations.
\item Matching all elements in $\{h_n^r\}$ with $\{h_{i,j}\}$, one obtains the pixels $(i,j)$ to which input $n$ contributes. Add this information to the lists $\left\{\{n\}_{i,j}\right\}$ accordingly.
\end{enumerate}
\end{enumerate}

The matching operation in step (3c) is where the lookup table approach comes into play. If $\{h_{i,j}\}$ is organized as a hash table that maps HEALPix indices to (lists of) target pixels $(i,j)$, then matching is just a lookup operation; if a HEALPix index from $\{h_n^r\}$ is in the hash table, the associated value is used to link $n$ to $(i,j)$.

\section{Cython implementation of the gridding algorithm: Cygrid}\label{sec:cygrid}

\subsection{Using Cython}

We provide a Cython-based implementation of the algorithm outlined above. Cython is a tool that facilitates writing Python-like code, which is then precompiled into C code. Cython can also be employed to wrap C/C++ libraries or source code to be used inside Python. In fact, we heavily use container classes (vectors and maps\footnote{In C++/STL a hash table is denoted as \textit{map}.}) from the C++ standard template library (STL) for cygrid .

\subsection{HEALPix functions}
Only a small subset of HEALPix routines is necessary for cygrid, mainly the functions to convert between world coordinates and HEALPix index. Instead of relying on the reference implementation of HEALPix, we reimplemented the required functionality in cygrid because it avoids the additional dependency on an external library. Furthermore, this allows us to add several tweaks for our purpose, for example, to the \texttt{query-disk} routine.

\subsection{A fast \texttt{query-disk} routine}\label{subsec:querydisk}

The largest benefit in terms of efficiency comes from performing the computation of the convolution only for pairs of input sample coordinates and output pixels that yield a significant weight factor. To find these pairs, we employ the \texttt{query-disk} function, which returns the indices of surrounding pixels for any HEALPix input pixel index, located inside a sphere of radius, $r_d$. The gridding kernel is then only calculated for samples within the queried region with $r_d$ chosen such that the values of the kernel function are practically zero outside the area enclosed by the sphere.

For each input world coordinate pair we first compute its HEALPix index. We only need to do the disk query once per HEALPix index instead of once per input world coordinate (number of involved HEALPix indices $\leq$ number of input coordinate pairs). Furthermore, we store the disk pixel lists in a hash table. Unless the input data is only sparsely sampled, this saves a fair amount of computing power because a hash table lookup is usually faster than the \texttt{query-disk} procedure.

To increase efficiency even further, we run the query-disk function only once per HEALPix ring, for the longitude $\alpha=180\degr$. By exploiting the regularity of the HEALPix grid, one can move the disks along longitude. For each disk pixel index (in RING scheme), $p$, we first calculate the ring number, $r$, and the number of HEALPix pixels in this ring, $n_r$.In the equatorial belt, $n_r$, is the same for each ring, while in the polar regions $n_r$ gets smaller toward the poles \citep[see][for details]{gorski05}. The (integer) shift, $\Delta p$, that needs to be applied to shift a pixel by an angle of $\Delta\theta$ can be calculated via
\begin{equation}
\Delta p = \frac{\Delta\theta}{2\pi} n_r\,,
\end{equation}
where $\Delta p$ is rounded to the nearest integer. If the index of the first pixel per ring is denoted as $s_r$, the new (shifted) pixel index is then given by
\begin{equation}
\tilde p =  s_r + \left[\left( p - s_r + \Delta p\right) \mod  n_r\right]\,.\label{eq:query_disk_shift}
\end{equation}

\subsection{Determine $\left\{\{n\}_{i,j}\right\}$}
We now briefly discuss how the lookup table approach works. The \texttt{grid} method of cygrid calls a function \texttt{calculate\_output\_pixels}, with the coordinates of the input samples as parameter. This function then runs the following two subroutines:

\begin{itemize}
\item \texttt{compute\_input\_output\_mapping:} This routine is straightforward. (1) Compute the HEALPix indices and ring indices for the provided input coordinates. (2) For each ring index, run the \texttt{query\_disk} routine (for $\alpha=180\degr$) and store in a hash table (or map, in C++ nomenclature). If an entry is already present in the hash table, the \texttt{query\_disk} is not be called again. For the typical case, this approach saves a huge amount of \texttt{query\_disk} calls, as the number of HEALPix rings involved is significantly smaller than the number of input coordinates. (3) For each input HEALPix index, now find the HEALPix indices of the surrounding disk, by shifting the $\alpha=180\degr$ disks as described in Section~\ref{subsec:querydisk}. As a result, we now have a hash table \textit{input\_output\_mapping}, which contains a list of target (WCS) pixels, associated to each input HEALPix index, for which the convolution has to be performed.
\item \texttt{compute\_output\_input\_mapping:} As a last step, we have to invert \textit{input\_output\_mapping}, i.e., calculate an \textit{output\_input\_mapping}, which contains the list of relevant input HEALPix indices for each target (WCS) pixel.
\end{itemize}

Many of the steps in the \texttt{calculate\_output\_pixels} function can also be parallelized, which accelerates processing when using multiple CPUs.

\subsection{Using cygrid: A minimal example}\label{subsec:cygrid_example}\lstset{
language=Python,
numbers=left,
numberfirstline=false,
xleftmargin=2.5em,
basicstyle=\ttfamily\small,
otherkeywords={self},
keywordstyle=\ttfamily\small\color{deepblue},
commentstyle=\color{mygrey},
emph={MyClass,__init__},
emphstyle=\ttfamily\small\color{deepred},
stringstyle=\color{deepgreen},
frame=tb,
showstringspaces=false,
}

\begin{figure}[!t]
\begin{lstlisting}[frame=,caption=Basic Cygrid usage.]  % Start your code-block

from astropy.io import fits
import cygrid

# read-in data
glon, glat, signal = get_data(...)

# define target FITS/WCS header
header = {
    'NAXIS': 3,
    'NAXIS1': 101,
    'NAXIS2': 101,
    'NAXIS3': 1024,
    'CTYPE1': 'GLON-SFL',
    'CTYPE2': 'GLAT-SFL',
    'CDELT1': -0.1,
    'CDELT2': 0.1,
    'CRPIX1': 51,
    'CRPIX2': 51,
    'CRVAL1': 12.345,
    'CRVAL2': 3.14,
    }

# prepare gridder
kernelsize_sigma = 0.2

kernel_type = 'gauss1d'
kernel_params = (kernelsize_sigma, )
kernel_support = 3 * kernelsize_sigma
hpx_maxres = kernelsize_sigma / 2

mygridder = cygrid.WcsGrid(header)
mygridder.set_kernel(
    kernel_type,
    kernel_params,
    kernel_support,
    hpx_maxres
    )

# do the gridding
mygridder.grid(glon, glat, signal)

# query result and store to disk
data_cube = mygridder.get_datacube()
fits.writeto(
    'example.fits',
    header=header, data=data_cube
    )
\end{lstlisting}
\end{figure}

In Listing~1 we show a minimal example of how a grid job could be programmed. After reading in the data in line~5 and the definition of a target-map FITS header (line~8) the user needs to create an instance of the \texttt{WcsGrid} class (line~31) and set an appropriate kernel (line~32).The \texttt{WcsGrid} class is an implementation of the \texttt{Cygrid} base class. The latter cannot be used directly, but only the derived classes work. Currently, we provide two implementations: \texttt{WcsGrid} and \texttt{SlGrid}. \texttt{WcsGrid} can be used to grid onto any of the WCSlib projections and must be provided with a FITS header during object construction (either as pyfits Header object or as Python dictionary). The \texttt{SlGrid} can be used to resample onto a list of coordinates (`sight lines').

In the call to the \texttt{set\_kernel} class method, we provide the desired kernel-function type and the associated parameters. Furthermore, the \textit{support-radius} and \textit{hpx-max-resolution} have to be set. The \textit{support-radius} defines the radius around each input coordinate sample out to which the convolution has to be performed. The chosen kernel function should have negligible contribution beyond the given radius. For example, for a radial-symmetric Gaussian kernel with standard deviation $\sigma$, the \textit{support-radius} could be chosen to lie between $3\sigma$ and $5\sigma$ depending on the desired accuracy. The other parameter \textit{hpx-max-resolution} defines the resolution of the internal HEALPix grid. All input coordinate samples and the output (WCS) map pixels (or sight lines) are internally associated with a HEALPix index to allow for optimal lookup table handling. For cygrid it is no issue if multiple input samples or output coordinate pairs share HEALPix indices, so in principle a very coarse HEALPix grid could be used. However, a very coarse HEALPix grid would mean that the \texttt{query-disk} function yield a frazzled disk. If the \textit{support-radius} is sufficiently large, this is not a problem, however, we recommend setting \textit{hpx-max-resolution} to about $\sigma/2$ if the disk radius is only set to $3\sigma$.

Also, the properties of the kernel and output grid should be compatible. To avoid aliasing effects, the output grid should be fine enough such that not only the input data but also the kernel function is fully sampled. Furthermore, unless the input data is not heavily oversampled, we advise choosing a kernel size that is not much smaller than the original resolution of the input data. Of course, the convolution-based gridding lowers the spatial resolution, $\sigma_\mathrm{data}$, of the gridded data somewhat. In typical scenarios, good results can be achieved with $\sigma_\mathrm{kernel}\approx0.5\sigma_\mathrm{data}$, such that the final resolution is
\begin{equation}
\sigma_\mathrm{data}^\mathrm{gridded} = \sqrt{\sigma_\mathrm{kernel}^2+\sigma_\mathrm{data}^2}\approx1.12\sigma_\mathrm{data}\,,\label{eq:convolreso}
\end{equation}
which is only slightly worse than the original resolution. A more detailed analysis on how to choose a well-suited kernel size is presented in Appendix~\ref{appsec:kerneldetails}.

Finally, after starting the grid job in line~40, we can retrieve the data cube (line~43) and write to a FITS file (line~44).

\subsection{Source code repository}\label{subsec:repository}

Cygrid is open-source software and is made available on the GitHub\footnote{\url{https://github.com/bwinkel/cygrid}} platform. Community contributions are welcome. We also prepared several Jupyter Notebooks\footnote{\url{http://jupyter.org/}} that demonstrate how to use cygrid in practice as follows:

\begin{itemize}
\item\textit{Minimal.} In this notebook, we show a minimal example of how to use cygrid, similar to Listing~1 but fully working. Moreover, we illustrate some applications of the \texttt{wcs} module from AstroPy\footnote{\url{http://www.astropy.org/}} in combination with cygrid.
\item\textit{Two fields.} This notebook explains how to grid data from one FITS-image coordinate projection to another, using \ion{H}{i} and far-infrared data.
\item\textit{Healpix allsky.} We demonstrate how to use the healpy library with cygrid and how one can grid full-sky data sets onto smaller FITS images.
\item\textit{Sightlines.} Cygrid is used to grid on individual lines of sight instead of WCS fields.
\end{itemize}
All these notebooks are also available on GitHub.

\section{Benchmarking cygrid}\label{sec:benchmarks}

\begin{figure}[!tp]
\centering%
\includegraphics[width=0.49\textwidth,viewport=16 42 523 392,clip=]{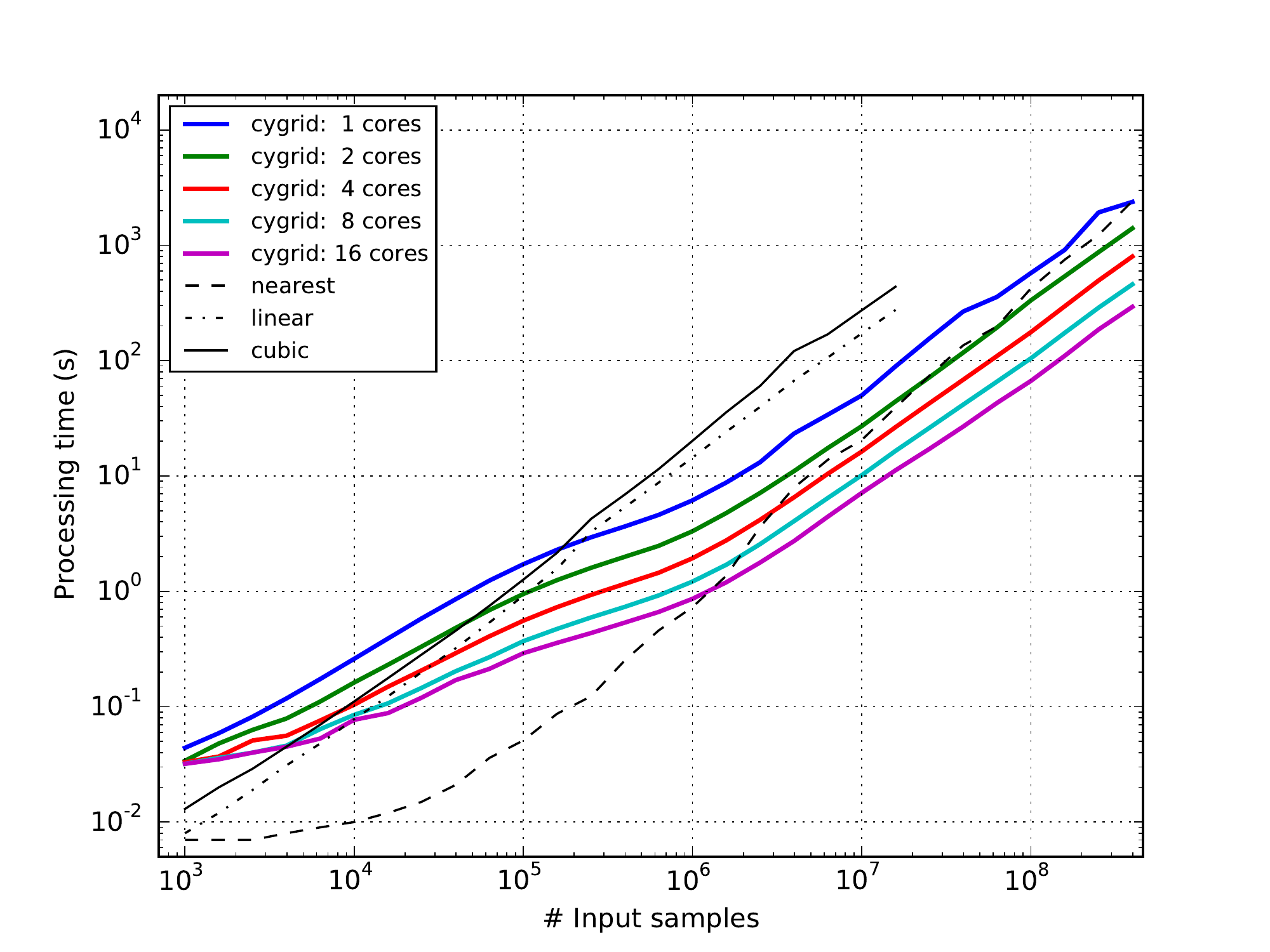}\\
\includegraphics[width=0.49\textwidth,viewport=16 4 523 391,clip=]{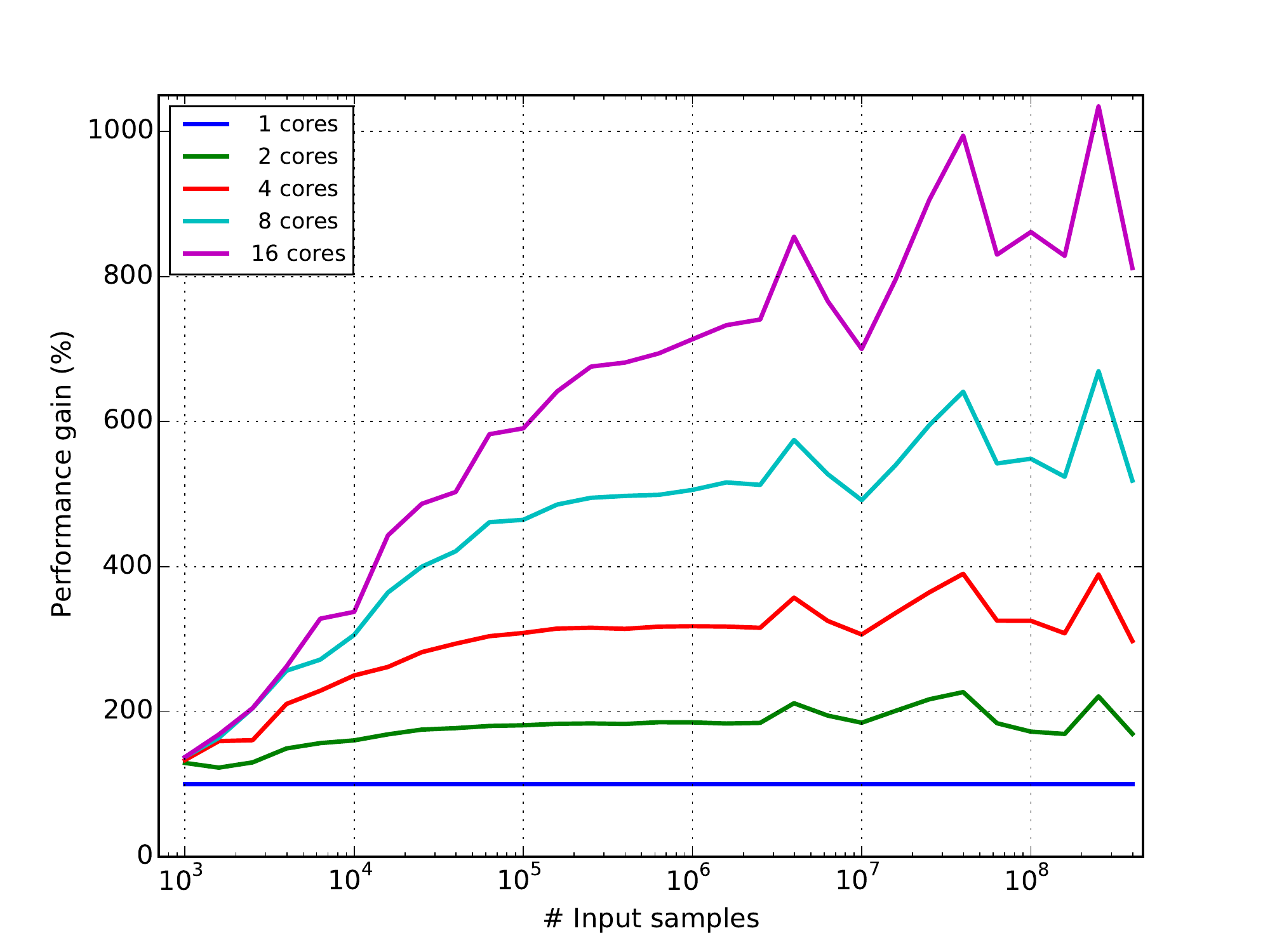}
\caption{\textit{Example~1:} Benchmark results for a small test program, which grids a different number of random samples to a target field of $5\degr\times5\degr$ size. The width of the gridding kernel was $\vartheta_\mathrm{fwhm}=300\arcsec$; the pixel size was $200\arcsec$.}%
\label{fig:benchmark_density}%
\end{figure}

In this short section, we analyze the performance of our cygrid implementation. For this, we produce random data; for a specified target field coordinate pairs (longitude and latitude) are sampled from a uniform distribution, and each is assigned a signal value (Gaussian noise). We then use cygrid to grid the data and measure the necessary computing time.

The result is displayed in Fig.~\ref{fig:benchmark_density} (top panel), which shows the processing times as a function of number of input samples for a field size of $5\degr\times5\degr$, a pixel size of $200\arcsec$, and a Gaussian kernel width of $\vartheta_\mathrm{fwhm}=300\arcsec$.

Next, we checked the quality of the parallelization. For this, we ran the test program with a different number of processor cores, from 2 to 16. In the lower panel of Fig.~\ref{fig:benchmark_density} we plot the relative performance gain of each run relative to single-threaded processing times. While a second core means almost 100\% speed-up, for a larger number of utilized cores the speed-up is less (per core). This reflects the nature of the gridding; it is not purely CPU bound, but is strongly dependent on the available memory bandwidth and cache performance. Still, on our workstation, for $10^8$ input samples, the 16-core processing time is just about 67~s.

For comparison, we carried out the same gridding task using the \texttt{scipy.interpolate.griddata} function\footnote{\url{http://docs.scipy.org/doc/scipy/reference/generated/scipy.interpolate.griddata.html}}, which provides nearest-neighbor and linear and cubic spline interpolation. The resulting runtimes are shown as black curves in Fig.~\ref{fig:benchmark_density}. For the spline interpolation, SciPy uses the QHull triangulation library\footnote{\url{http://www.qhull.org/}}, which apparently limits the number of samples to about $16.7$ million. Therefore, only the nearest-neighbor interpolation was carried out up to the maximum of $10^8$ samples. For many input samples, the spline interpolation is about an order of magnitude slower than cygrid, while the much simpler nearest neighbor algorithm is about twice as fast (compared to a single-threaded cygrid). The \texttt{griddata} function is not the best choice for data gridding, but is aimed at resampling data sets. For cases with very dense sampling of the input data (i.e., when the gridding operation is effectively downsampling a data set) the RMS noise in the target map is not decreased. We demonstrate this effect in the first of the Jupyter notebooks; see Section~\ref{subsec:repository}. Therefore, we did the comparison with SciPy with respect to the runtimes but not the resulting images.

\begin{figure}[!tp]
\centering%
\includegraphics[width=0.49\textwidth,viewport=20 4 523 392,clip=]{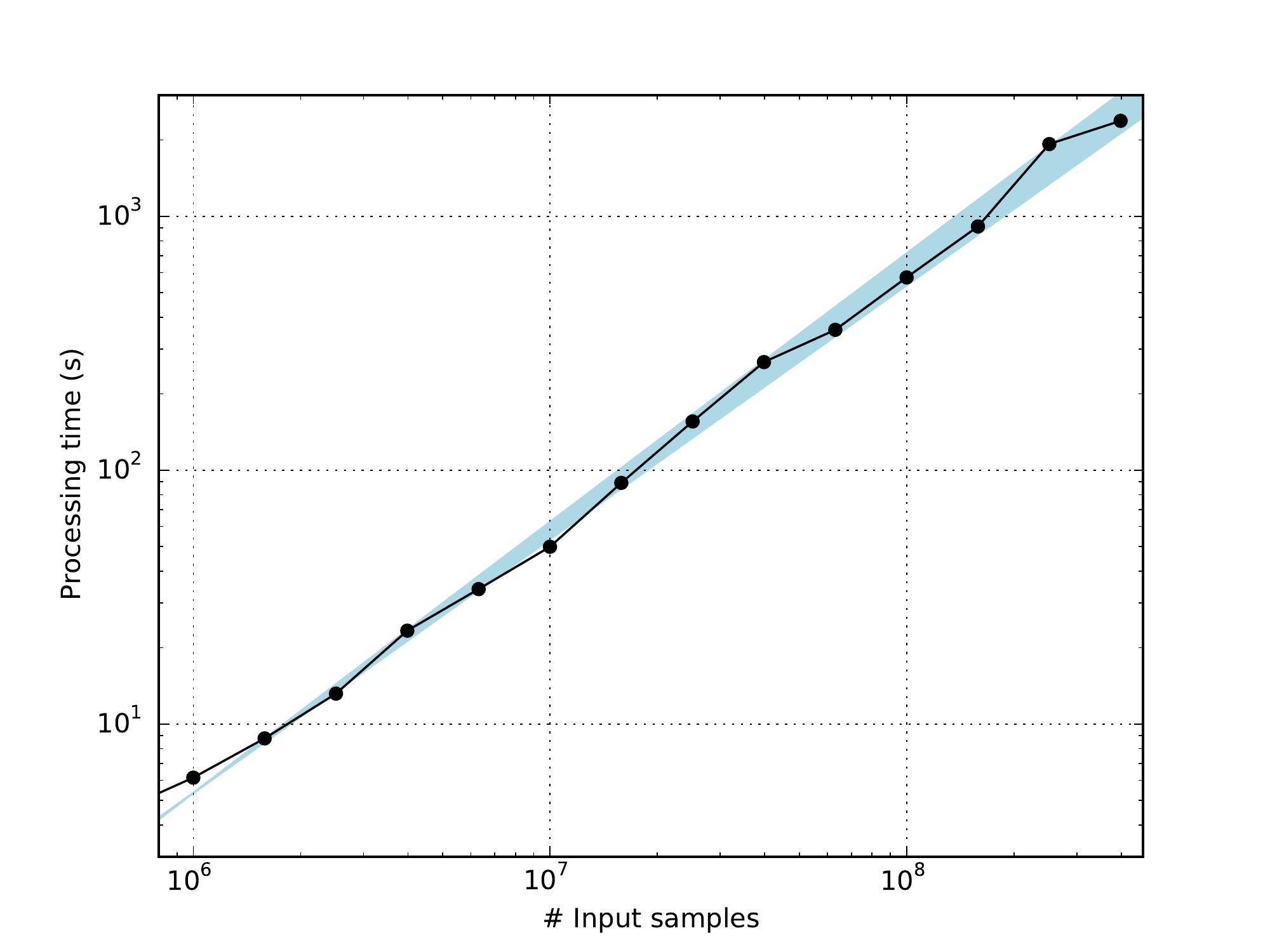}
\caption{Asymptotic behavior of cygrid's runtimes for constant field size. The time complexity (black solid line) appears to lie within $O(n)$ and $O(n\log n)$ (blue shaded area), where $n$ is the number of input samples.}%
\label{fig:benchmark_complexity}%
\end{figure}

The asymptotic behavior in Fig.~\ref{fig:benchmark_density} suggests a computational complexity, which is close to linear, i.e., $O(n)$. To visualize this better, we plot in Fig.~\ref{fig:benchmark_complexity} the 1-core curve (solid black line) from the example along with a blue shaded area that indicates the $O(n)$ and $O(n\log n)$ curve complexities. The latter are scaled such that they yield the same processing time for $10^6$ samples and enclose a solid curve over most of the asymptotic regime. The measured processing times are somewhat noisy, but it can be seen that the behavior is indeed compatible with $O(n)$ and $O(n\log n)$ behavior. The fact that the complexity is a little worse than $O(n)$ is probably caused by overheads in the hash table lookup for a huge number of entries.

\begin{figure}[!tp]
\centering%
\includegraphics[width=0.49\textwidth,viewport=16 42 523 392,clip=]{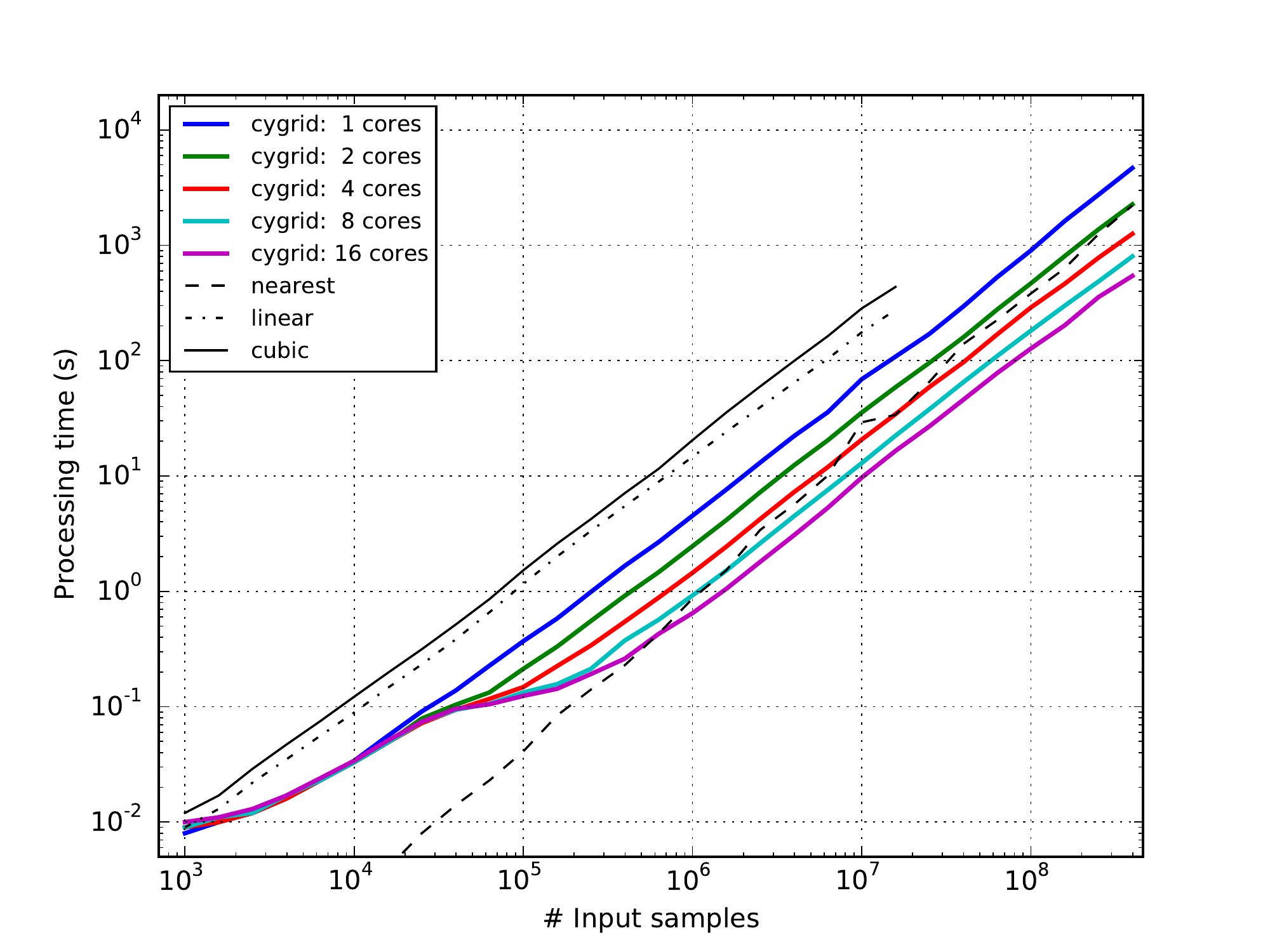}\\
\includegraphics[width=0.49\textwidth,viewport=16 4 523 391,clip=]{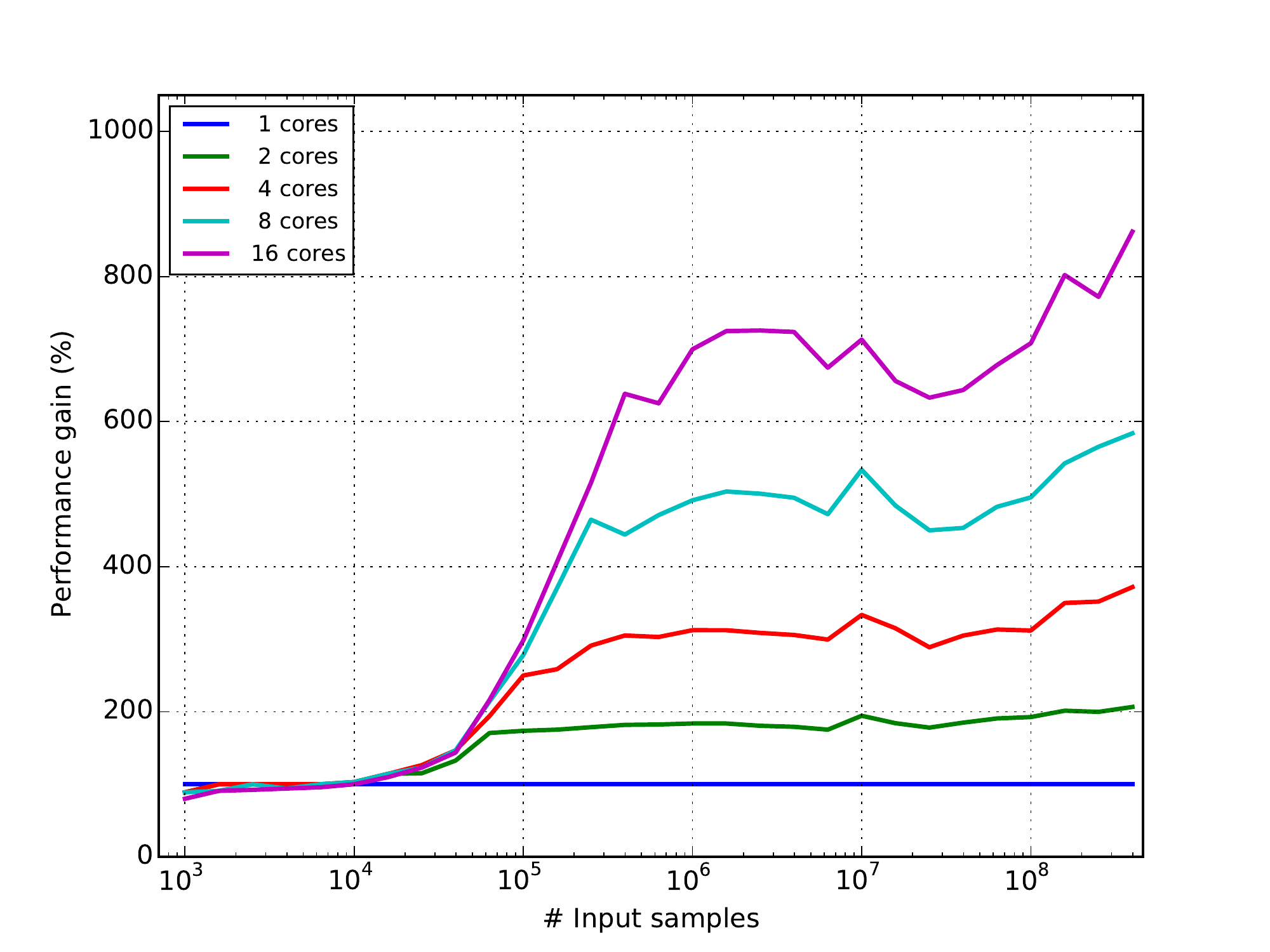}
\caption{\textit{Example~2:} As in Fig.~\ref{fig:benchmark_density} but for varying field sizes of $0.1\degr$ to $60\degr$ edge length. The spatial pixel density was kept constant at 100\,000 samples per square degree.}\label{fig:benchmark_fieldsize}%
\end{figure}

We carried out a second test where we kept the spatial sample density constant and varied the field size from about $0.1\degr$ to $60\degr$ (using a spatial density that yields the same number of samples as in the first test for better comparison). In Fig.~\ref{fig:benchmark_fieldsize} we show the results in the same way as for Fig.~\ref{fig:benchmark_density}. The pixel size and kernel width was the same as for the first test. Compared to the first benchmark, cygrid is about a factor of two slower. We attribute this to the increased number of \texttt{query-disk} calls and to the much larger disk hash table, which causes slower lookup times. We compare the runtimes with SciPy's \texttt{griddata} function, and cygrid is still competitive (factor of 2 slower with 1-core) or much faster (multiple cores).

\section{Summary}\label{sec:summary}

We have presented cygrid, a fast and flexible data gridding module for the Python programming language. Cygrid implements a convolution-based algorithm and, to maximize its performance, makes use of hash tables, a fast query-disk routine, and the Cython-extension.

Cygrid can be used to resample onto any of the FITS/WCS projections to easily create astronomical maps or data cubes. It is also possible to grid to a list of coordinates (sight lines), which can  be used, for example, to produce maps on the HEALPix grid.

It was demonstrated that our cygrid implementation benefits from multicore CPUs (with relatively good scaling behavior) and easily outperforms other widely used gridding algorithms, such as the linear and cubic spline interpolation, which ship with SciPy. Cygrid's runtimes to grid a 2D map lie between $O(n)$ and $O(n\log n)$ complexity, i.e., the overheads compared to a linear behavior, are very moderate. For data cubes the impact of the overheads is even smaller because all samples of one spectrum are multiplied with the same weight factor.

\begin{acknowledgements}

We are grateful to Alexander Kraus for carefully proofreading the manuscript and for his valuable comments.

Some of the results in this paper were derived using the HEALPix \citep{gorski05} package. Cygrid was developed in the framework of the Effelsberg-Bonn \ion{H}{i} Survey (EBHIS) project. EBHIS is based on observations with the 100-m telescope of the MPIfR (Max-Planck-Institut für Radioastronomie) at Effelsberg. The authors thank the Deutsche Forschungsgemeinschaft (DFG) for support under grant numbers KE757/7-1, KE757/7-2, KE757/7-3, and KE757/9-1. B.W. was partially funded by the International Max Planck Research School for Astronomy and Astrophysics at the Universities of Bonn and Cologne (IMPRS Bonn/Cologne). L.F. was also a member of IMPRS Bonn/Cologne. D.L. is a member of the Bonn--Cologne Graduate School of Physics and Astronomy (BCGS).

We would like to express our gratitude to the developers of the many C/C++ and Python libraries, which are made available as open-source software and we used: most importantly, NumPy \citep{NumPy} and SciPy \citep{SciPy}, Cython \citep{Cython}, and Astropy \citep{Astropy}. Figures were prepared using matplotlib \citep{Matplotlib} and in part using the Kapteyn Package \citep{KapteynPackage}.

\end{acknowledgements}

\bibliographystyle{aa} 
\bibliography{references} 

\appendix

\section{The gridding kernel revisited}\label{appsec:kerneldetails}

To use cygrid, one needs to choose a gridding kernel. The choice of kernel type is usually application dependent, but the kernel size and kernel support radius must be chosen well. In the following, several aspects are discussed to guide the potential user. We restrict the analysis to the case of a (spherical-symmetric) Gaussian kernel for the sake of simplicity .

\subsection{Image resolution}
Any convolution-based gridding algorithm degrades the resolution of the imaged signal at least somewhat -- with the exception of the $\sin(ax)/ax$ kernel, which can reconstruct the original input signal if computed with (effectively) infinite support. From Eq.~(\ref{eq:convolreso}) it is obvious that the smaller the convolution kernel, the better the resolution after gridding. A very subtle effect, however, lies in the potential different correlation lengths of the input signal and measurement noise. In many astronomical applications, the instrument itself adds noise to the measured signal, for example, produced by a CCD detector or due to the thermal noise component in a radio receiver. There are even cases, when the noise is effectively independent in each sample (e.g., for a single-dish radio  observation), while the desired astronomical signal is convolved with the beam or point spread function (PSF) of the instrument and as such is correlated in the spatial domain.

\begin{figure}[!t]
\centering%
\includegraphics[width=0.49\textwidth,viewport=40 20 630 320,clip=]{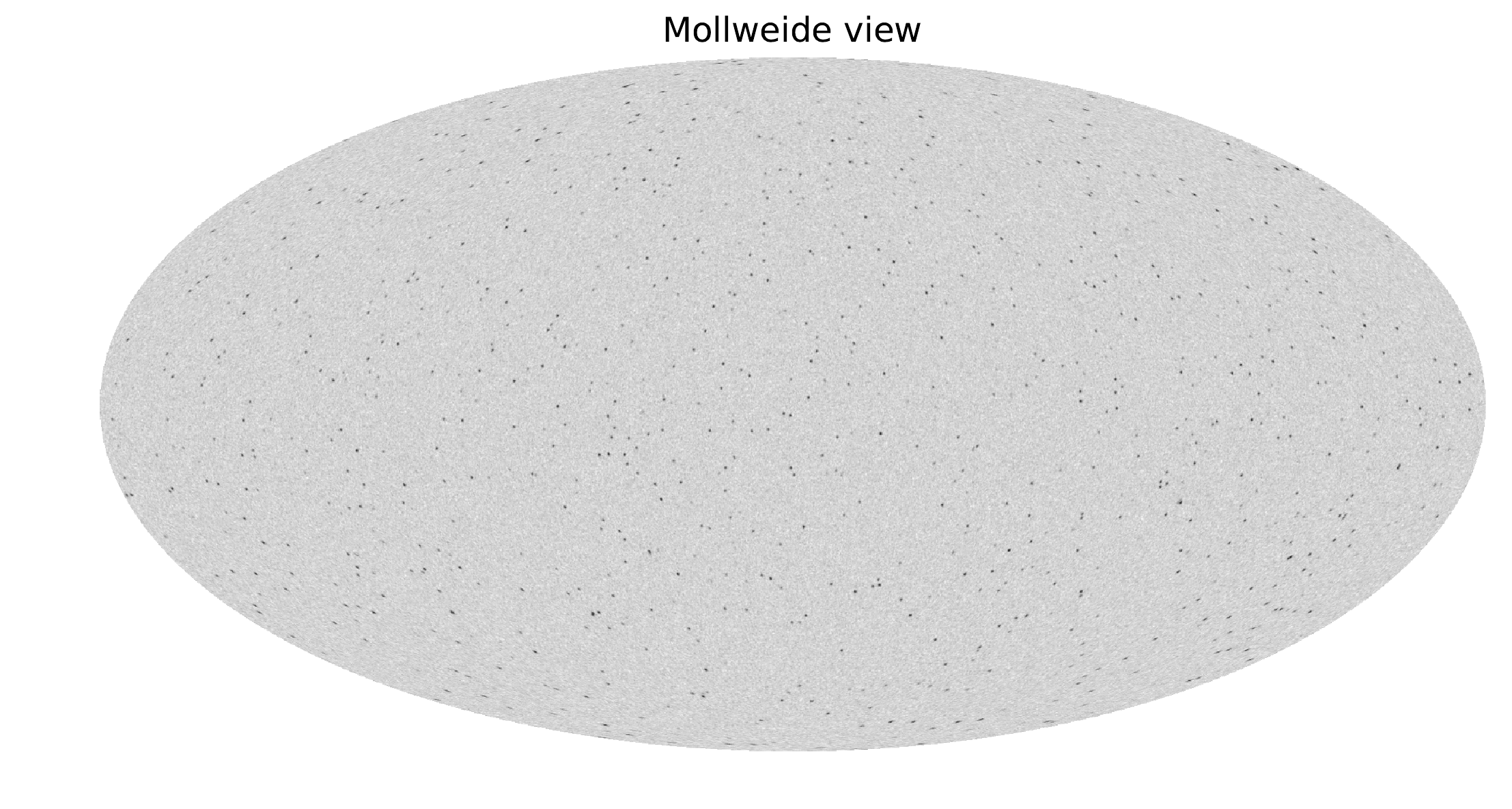}
\caption{A simulation is used to study the impact of the gridding on the power spectral densities of the input signal and noise. We randomly placed 1024 point sources with different intensities on the sphere. To generate the `measured' input signal, the true point-source signal is convolved with a PSF of $\vartheta_\mathrm{psf}=27\farcm5$ and sampled on about 50 million random coordinate pairs. Independent Gaussian noise is added to each resulting signal sample. Using a Gaussian-shaped gridding kernel with $\vartheta_\mathrm{kernel}=13\farcm7$ the input signal and noise are then imaged to the nside=512 HEALPix grid.}%
\label{fig:hpx_image}%
\end{figure}

To visualize this we run a simulation, gridding artificial data to an all-sky map. To avoid map projection effects, we grid to the HEALPix nside=512 grid pixels, $(l_\mathrm{hpx}, b_\mathrm{hpx})$, (pixel size: $\vartheta_\mathrm{hpx}=6\farcm87$) using cygrid's sightline gridder (\texttt{SlGrid}). The artificial signal is comprised of 1024 point sources that are randomly distributed on the sphere with different intensities. For a realistic scenario, the effect of an observing instrument needs to be considered. Therefore, we also randomly sample about 50 million\footnote{This is 15 times the number HEALPix grid pixels, such that on average 15 samples contribute to each pixel in the gridded map.} observing positions $(l_\mathrm{in}, b_\mathrm{in})$ that are uniformly distributed over the sphere. A Gaussian-type PSF with $\vartheta_\mathrm{psf}=4\vartheta_\mathrm{hpx}=27\farcm5$ (FWHM; { $\vartheta_\mathrm{fwhm}=\sqrt{8\ln2}\sigma\approx2.35\sigma$}) is then applied to calculate the `measured' signal, $s_\mathrm{in}$, on each of the input coordinate samples $(l_\mathrm{in}, b_\mathrm{in})$. For each of the $s_\mathrm{in}$ samples we draw a noise sample, $n_\mathrm{in}$, from a normal distribution. The simulated data points, $s_\mathrm{in}+n_\mathrm{in}$, are then gridded using a kernel of $\vartheta_\mathrm{kernel}=\vartheta_\mathrm{psf}/2=13\farcm7$ in size (FWHM). The resulting image is shown in Fig.~\ref{fig:hpx_image}.

\begin{figure}[!t]
\centering%
\includegraphics[width=0.49\textwidth,viewport=2 4 430 416,clip=]{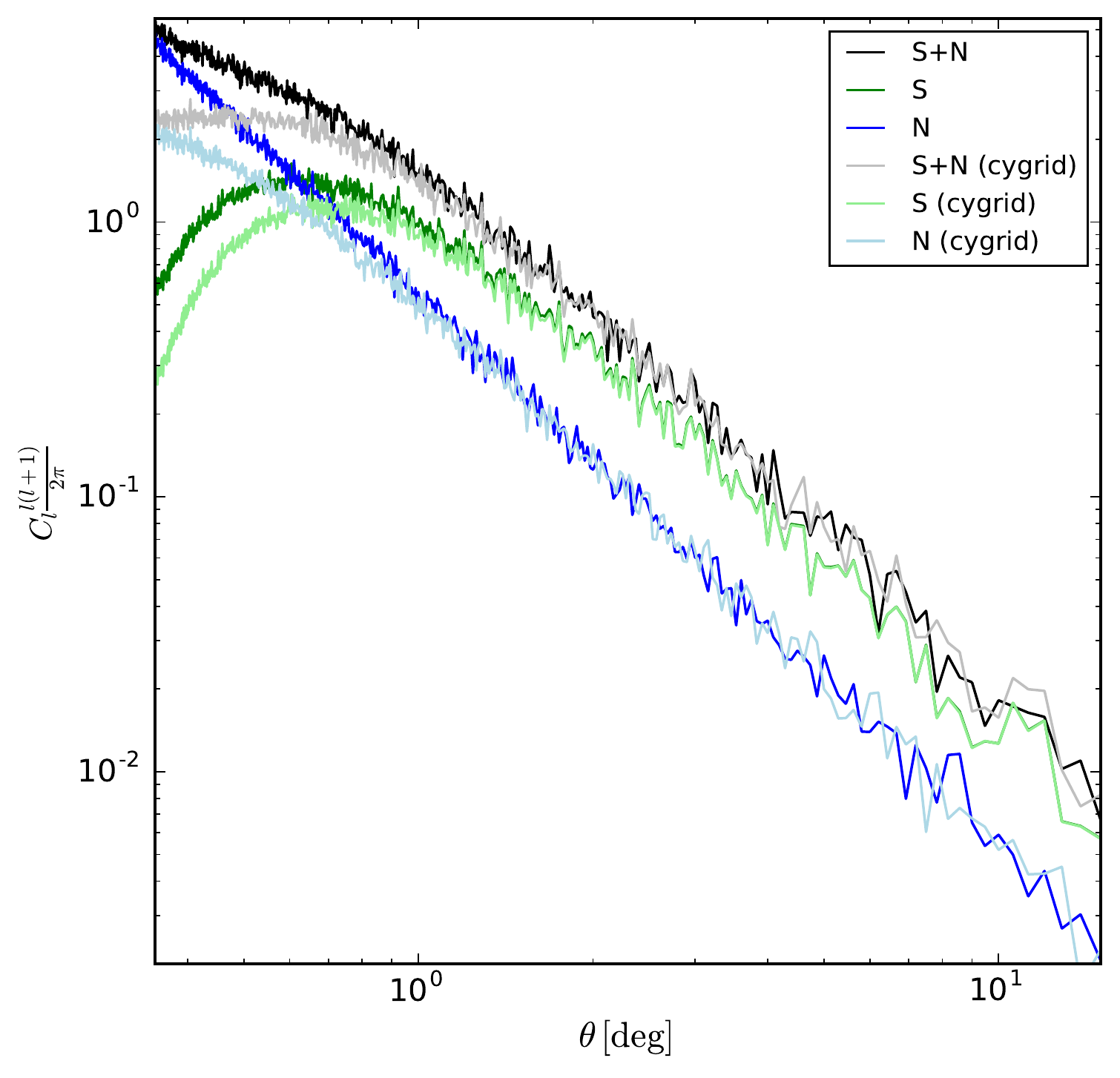}
\caption{Power spectral densities as calculated using the healpy routine \texttt{anafast} for the simulations. See text for details.}%
\label{fig:psd}%
\end{figure}

Using the healpy routine \texttt{anafast} one can easily calculate the power spectral density (PSD) of such a signal on the sphere. In Fig.~\ref{fig:psd} the gray solid curve indicates the PSD returned by \texttt{anafast} for the signal plus noise after gridding. We also processed the input signal and noise separately (light green and light blue solid lines, respectively).

To study the effect of the gridding itself, a comparison with the true signal is helpful. Therefore, we calculate the PSF-convolved representation of the true signal on the HEALPix pixel centers directly (green solid line in Fig.~\ref{fig:psd}). As before, independent noise samples are drawn (blue solid line). For consistency with the gridded noise values, a scaling needs to be applied to account for the larger number of samples in the gridding case. The combined theoretical signal plus noise is shown with a black solid line.

Both the gridded signal and gridded noise have less power on the small angular scales, compared to the true signal and noise spectral densities. This is because the convolution smooths the data on small scales. The true input signal (but not the noise) already shows a drop-off in the spectral power toward small angular scales because the point sources were convolved with a Gaussian PSF. One important aspect to keep in mind is that, as shown, the resulting spatial resolution (or rather correlation-length) of the noise after gridding can be different from the resulting spatial resolution of the signal. Also, cygrid conserves the total flux density in the image, which, for example,  means that the amplitude of a (Gaussian) peak decreases while its width increases during convolution with the kernel.

\subsection{Aliasing and choice of kernel size}
\begin{figure*}[!p]
\centering%
\includegraphics[width=0.98\textwidth,viewport=5 10 855 1070,clip=]{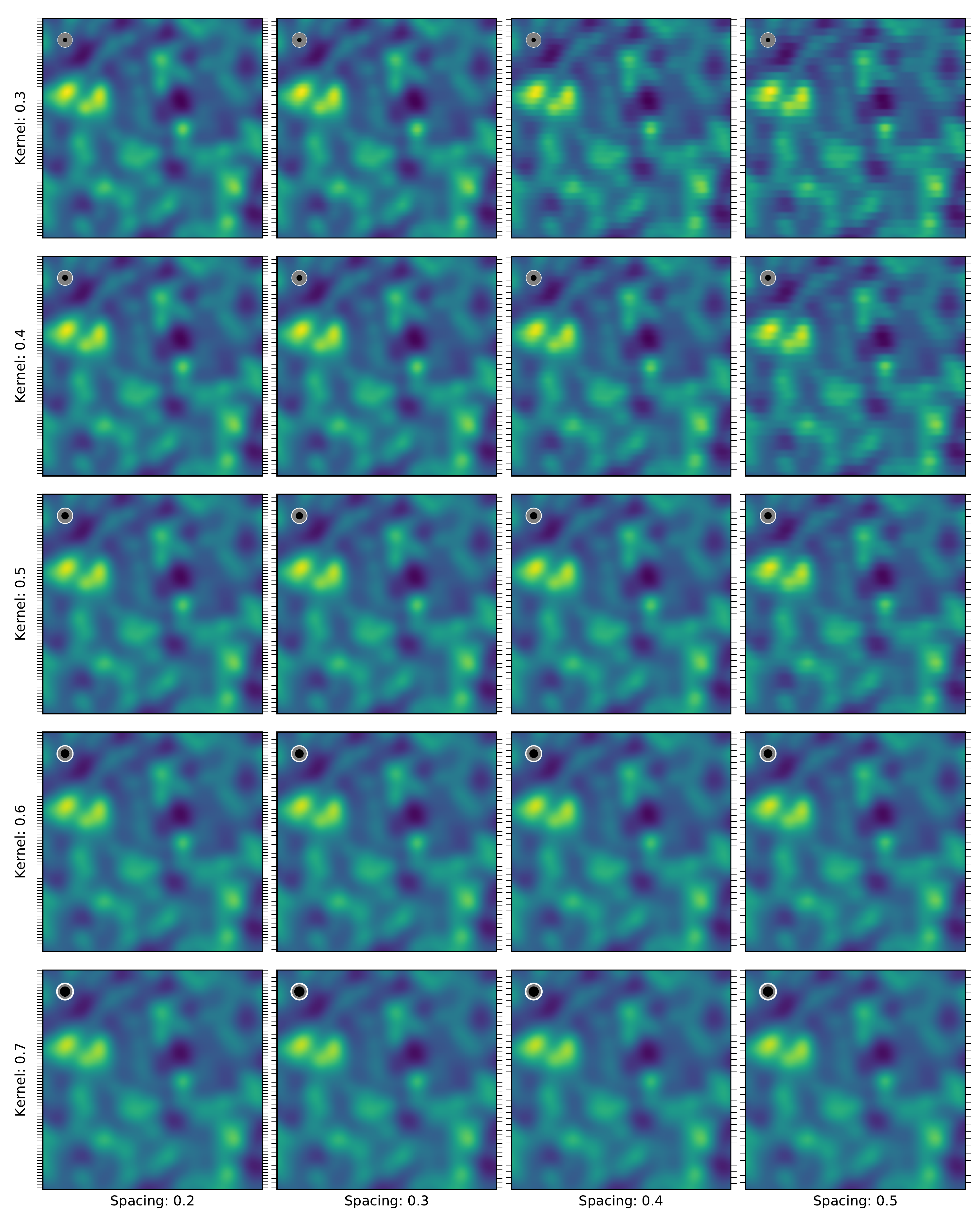}
\caption{Impact of different sampling densities and kernel sizes on the accuracy of the gridding algorithm. Input samples are distributed on horizontal stripes (scan lines), separated with the given spacing (in units of the beam size). The locations of the stripes are indicated by the tick marks on the vertical axes. Various kernel sizes (also in units of beam size) were applied to image the artificial signal. In the top left of each panel the kernel size (black circle), beam size (gray circle), and resulting (theoretical) image resolution (white circle) are visualized. See text for further details.}%
\label{fig:aliasing_5_20_image}%
\end{figure*}

The above discussion does not allow us to come to a conclusion on the choice of an optimal kernel size. Why can we not use an arbitrarily small kernel, for example, which would avoid the degradation of the spatial resolution? In fact, the necessary kernel size is entirely determined by the sampling density of the input data point and the required accuracy. To demonstrate this, we set up a different simulation. The input coordinates are now placed on horizontal stripes (scan lines), with very dense sampling along the stripe direction, but with a certain spacing between the stripes. Such a scenario is common for mapping observations with a single-dish radio telescope with just one feed horn. It also effectively converts the question about the proper kernel size to a one-dimensional problem, which is easier to analyze. As before, we construct an artificial signal from point sources, convolved with $\vartheta_\mathrm{psf}$, but here we use a much higher source density to obtain a quasi-continuous field. The noise component is omitted this time because we want to analyze potentially small effects in the image reconstruction. We use four different stripe spacings, $d_\mathrm{space}=[0.2,\,0.3,\,0.4,\,0.5]\times\vartheta_\mathrm{psf}$ and grid each of the resulting input signals with five different kernel sizes, $\vartheta_\mathrm{kernel}=[0.3,\,0.4,\,0.5,\,0.6,\,0.7]\times\vartheta_\mathrm{psf}$. The field size is $1.5\degr\times1.5\degr$ and $\vartheta_\mathrm{psf}=0.1\degr$ (FWHM), but as the spacing and kernel sizes are relative to the beam sizes this is unimportant for the results. We chose the pixel size of the target grid to be $\vartheta_\mathrm{pix}=\vartheta_\mathrm{psf}/20$, which is heavily oversampled, but useful for revealing all of the potential effects. For the same reason, the support radius was made large with $5\sigma_\mathrm{kernel}$. Figure~\ref{fig:aliasing_5_20_image} shows the resulting maps. In the top left of each panel the kernel size (FWHM, black circle), the beam size (FWHM, gray circle), and the resulting (theoretical) image resolution (FWHM, white circle) are visualized. The tick marks on the vertical axes depict the position of the horizontal stripes on which the input samples are located.

\begin{figure*}[!p]
\centering%
\includegraphics[width=0.98\textwidth,viewport=5 10 855 1070,clip=]{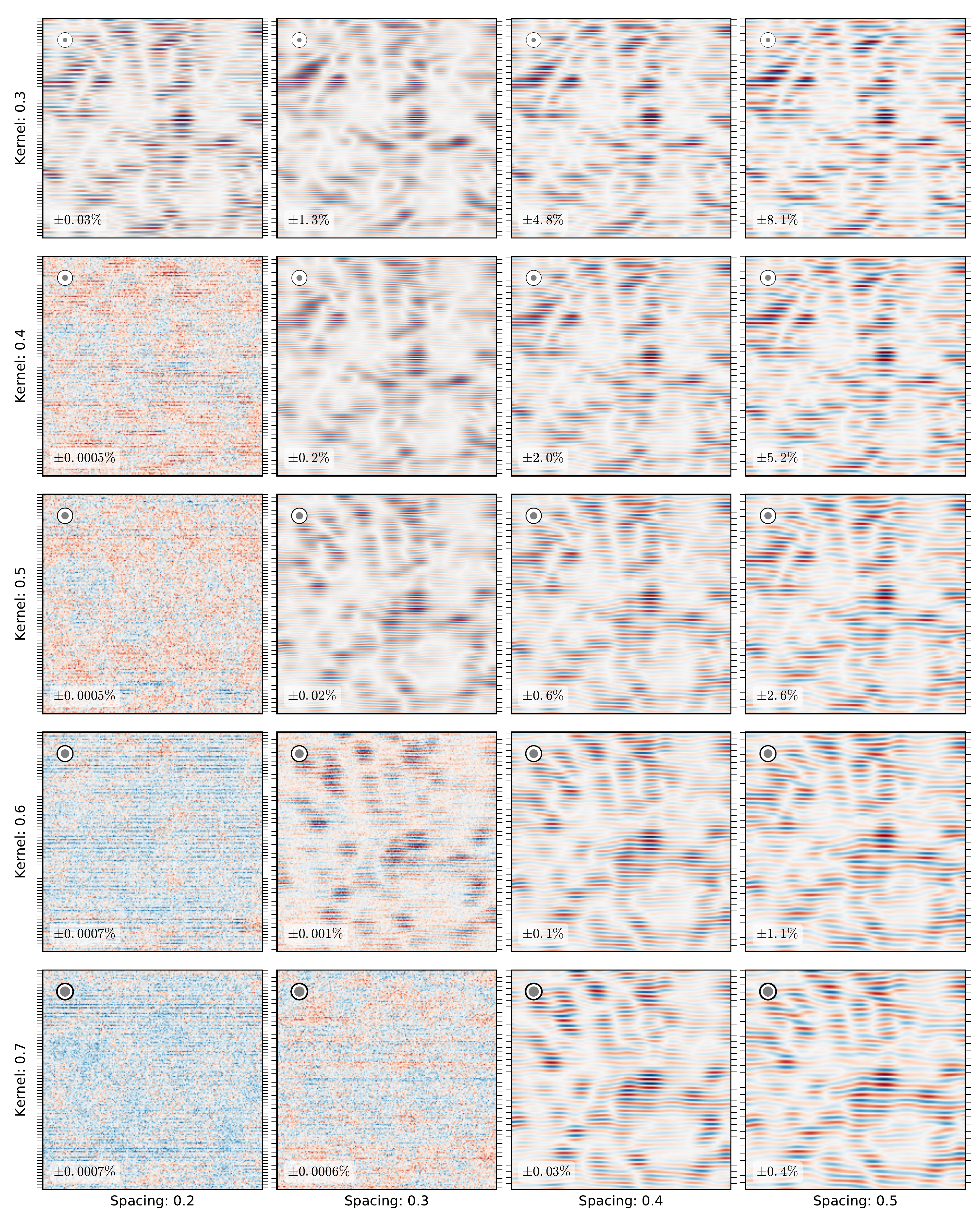}
\caption{As Fig.~\ref{fig:aliasing_5_20_image} but showing the relative deviation from the theoretical signal (convolved with the gridding kernel). See text for further details.}%
\label{fig:aliasing_5_20_diff}%
\end{figure*}

In the top right panel of Fig.~\ref{fig:aliasing_5_20_image} ($d_\mathrm{space}=0.5\vartheta_\mathrm{psf}$, $\vartheta_\mathrm{kernel}=0.3\vartheta_\mathrm{psf}$) block-like artifacts are clearly visible, which hint at a problem with the imaging (so-called aliasing). For a more detailed analysis, we show in Fig.~\ref{fig:aliasing_5_20_diff} the relative deviation of the gridded image from the input signal. The latter was also convolved with the gridding kernel, otherwise the different spatial resolutions would not facilitate a meaningful comparison. As the deviation maps have extremely different scales, we multiplied each panel with a scale factor. The number in the lower left of each panel indicates the minimal and maximal deviation for each case associated with the dark blue and red colors in the image, respectively.

\begin{figure*}[!p]
\centering%
\includegraphics[width=0.98\textwidth,viewport=5 10 855 1070,clip=]{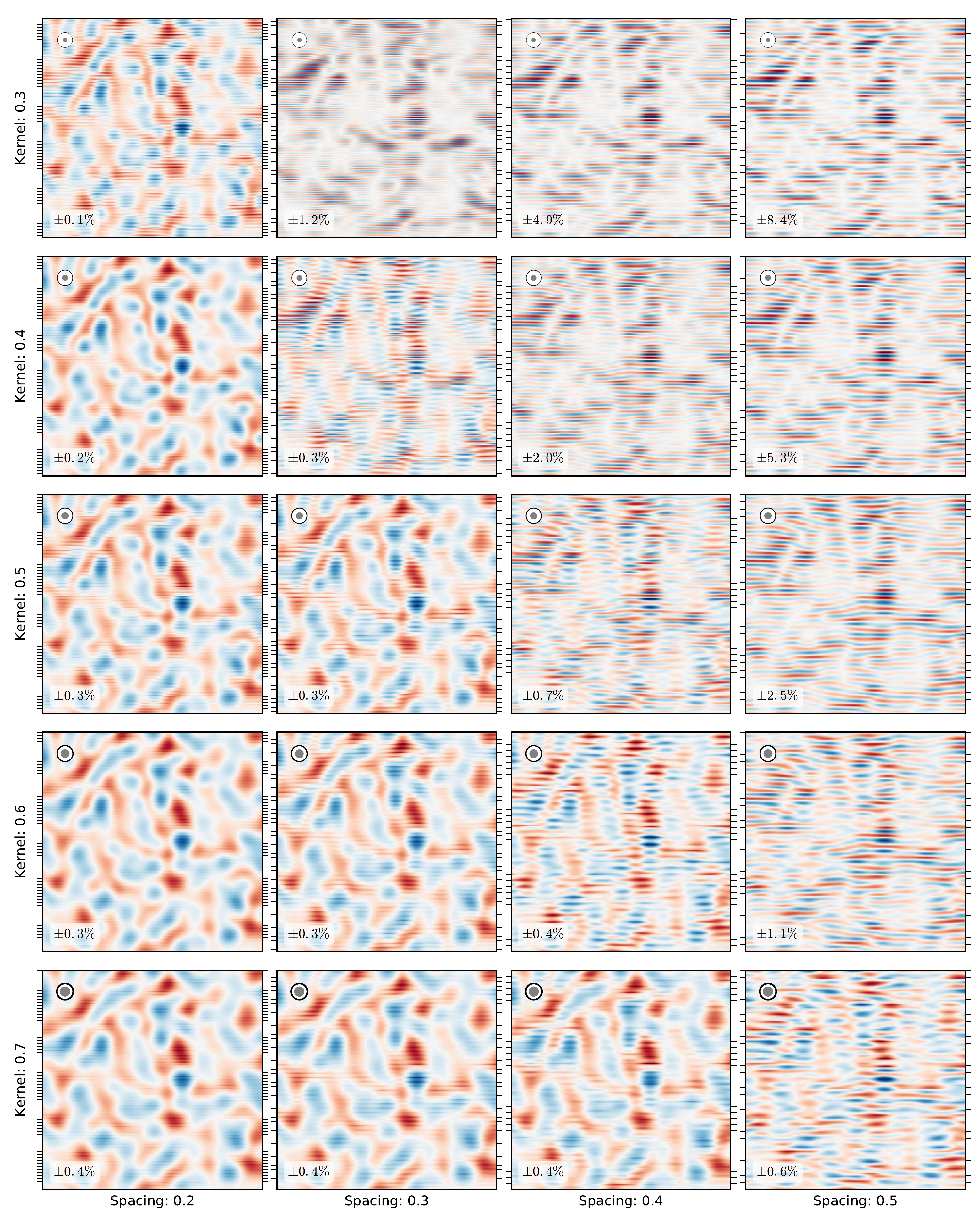}
\caption{As Fig.~\ref{fig:aliasing_5_20_diff} but with a smaller kernel support radius of $3\sigma_{kernel}$. See text for further details.}%
\label{fig:aliasing_3_20_diff}%
\end{figure*}

For the demonstrated setup with separated scan lines we could now choose the kernel size as a compromise between accuracy and acceptable degradation of image resolution. Of course, if a very different data point sampling or different kernel type was used, a similar analysis should be employed to find well-suited kernel parameters. As a rule of thumb, a kernel size of 0.5$-$0.6$\vartheta_\mathrm{psf}$ probably provides results accurate to better than 1$-$2\% even for relatively bad sampling densities. We also note, however, that the choice of the kernel support radius has an impact on the maximal accuracy. To demonstrate this, we repeated the simulation with a smaller support radius of $3\sigma_\mathrm{kernel}$; see Fig.~\ref{fig:aliasing_3_20_diff}. Comparing the two figures reveals that the support radius mainly impacts the high-accuracy cases, while the low-precision cases are not affected much.

\end{document}